\documentclass[prl,aps,twocolumn]{revtex4}

\usepackage{graphics}

\begin{document}

\title{Secure sharing of random bits over the Internet}

\author{Geraldo A. Barbosa$^*$}

%\affiliation{Northwestern University, Department of Electrical Engineering and Computer Science, 2145 N. Sheridan Road,
%Evanston, IL 60208-3118 }

\affiliation{QuantaSec -- Research in Quantum Cryptography Ltd., 1558 Portugal Ave., Belo
Horizonte MG 31550-000 Brazil. }

\date{17 May 2007}
\begin{abstract}

Although one-time pad encrypted files can be sent through Internet channels, the need for renewing shared secret keys have
made this method unpractical. This work presents a scheme to turn practical the fast sharing of random keys over arbitrary
Internet channels. Starting with a shared secret key sequence of length $K_0$ the users end up with a secure new sequence $K
\gg K_0$. Using these sequences for posteriori message encryption the legitimate users have absolute security control
without the need for third parties. Additionally, the security level does not depend on the unproven difficulty of factoring
numbers in primes. In the proposed scheme a fast optical random source generates random bits and noise for key renewals. The
transmitted signals are {\em recorded} signals that carries both the random binary signals to be exchanged and {\em
physical} noise that cannot be eliminated by the attacker. These signals allow amplification over the Internet network
without degrading security. The proposed system is also secure against a-posteriori known-plaintext attack on the key.
Information-theoretic analysis is presented and bounds for secure operation are quantitatively determined.

%{\bf KeyWords:} {\em Cryptography, Physical
%Noise, Internet, Secret Key, No Third-Party, fast communication,
%amplification allowed}

\end{abstract}

\maketitle

\newcommand{\be}{\begin{equation}}
\newcommand{\ee}{\end{equation}}
\newcommand{\bea}{\begin{eqnarray}}
\newcommand{\eea}{\end{eqnarray}}

\section{Introduction}

Unconditionally secure one-time pad encryption \cite{vernam} has not find wide applicability in modern communications. The
difficult for users to share long streams of secret keys beforehand has been an unsurmountable barrier preventing widespread
use of one-time pad systems. Even beginning with a start sequence of shared secret keys, no practical amplification method
to obtain new key sequences or key ``refreshing'' is available. This work proposes a solution for this problem where
transmitted random bit sequences are protected {\em physical} noise. Secure operational bounds are imposed by the noise
level used and a controlled {\em number} of bits transmitted. Posterior data encryption is done by $X$-or bit-by-bit the
shared random bits with the message bits as in one-time pad encryption.

The method involves binary random sequences physically created (key sequence) ${\bf X}=x_1,x_2,...$ to be transmitted
between the legitimate users A to B, and physical noise sequences ${\bf n}=n_1,n_2,...$ that are not controlled or reducible
by any means and will be physically superposed on ${\bf X}$. A and B share a starting secret key ${\bf K}_0$. Also essential
for the method is the use of physical {\em non-orthogonal} bases specified by shared keys ${\bf K}$ to encode the message
bits ${\bf X}$, before the deterministic ``recording'' of the noisy signals described by ${\bf Y}$.  The transmitted signals
are open to the attacker but improving signal resolution is impossible over the deterministic patterns. This simplifies the
security considerations, distinctly from cases where the attacker have access to physical signals in the channel and may
constantly improve its technology for signal resolution by performing homodyne, heterodyne and any other measurement.

It will be assumed that (statistical) physical noise ${\bf n}$ can be added to a message bit sequence ${\bf X}$ according to
some rule $f_j(x_j,n_j)$ giving ${\bf Y}=f_1(x_1,n_1),f_2(x_2,n_2),...\:$ (Whenever only binary physical signals are
implied, use of $f_j(x_j,n_j)$ will represent $f_j \equiv \oplus$ ($=$addition mod2)). When analog physical signals are made
discrete by analog-to-digital converters,  a sum of a binary signal onto a discrete set will be assumed or other convenient
rounding of the signal. $f_j$ describe operations on non-orthogonal bases specified by shared keys  ${\bf K}$ as will be
explained. The addition process is performed at the emitter station and ${\bf Y}$ becomes a binary file carrying random bits
and the recorded noise. ${\bf Y}$ is sent from user A to user B (or from B to A) through an insecure channel and can be
tapped with perfection (any copy will be identical to each other) by the attacker. The amount of noise is assumed high and
such that without any knowledge beyond ${\bf Y}$, {\em neither} B (or A) {\em or} an attacker E could extract the sequence
${\bf X}$ with a probability $P$ better than $P\simeq(1/2)^N$, where $N$ is the number of bits transmitted. The conditional
entropy for these randomized signals satisfy
\begin{eqnarray}
H({\bf Y}|{\bf K}_0,{\bf X})\neq 0\:\:,
\end{eqnarray}
 that guarantees that even knowing the message ${\bf X}$ and the key ${\bf K}_0$, the transmitted sequence ${\bf
Y}$ is not unique. This emphasizes the uncontrollable character of the physical noise present at each signal generation.

Assuming that A and B share some knowledge beforehand (the key ${\bf K}_0$ or ${\bf K}$), the amount of information between
A (or B) and E differs. This information {\em asymmetry} can be expressed by
\begin{eqnarray}
H_{AB}({\bf X}|{\bf Y},{\bf K})(\approx 0)\ll H_E({\bf X}|{\bf Y})(\approx H({\bf X}))\:.
\end{eqnarray}

The mutual information reflects that  E has much less information on $H({\bf X})$ than A and B:
\begin{eqnarray}
I_{AB}({\bf X};{\bf Y},{\bf K})&=&H({\bf X})- H_{AB}({\bf X}|{\bf Y},{\bf K})\approx H({\bf X}),\\
 I_{E}({\bf X};{\bf Y})&=&H({\bf X})-
H_{E}({\bf X}|{\bf Y})\approx 0\:.
\end{eqnarray}

 This asymmetry    will be used by A and B to
share secure information over the Internet. It will be shown that if A and B start sharing a secret key sequence ${\bf K}_0$
they end up with a secure new key sequence ${\bf K} \gg {\bf K}_0$. Within bounds to be demonstrated, this makes bit-by-bit
encryption (as a one-time pad) practical for fast Internet communications (data, image or sound). It should be emphasized
that being secure does not imply that ${\bf K}_0$ can be open to the attacker after transmission. All keys, ${\bf K}_0$ and
${\bf K}$,  have to be kept secret as long as messages have to be protected, as in one-time pad communication.

The system gives users A and B direct control to guarantee secure communication without use of third parties or
certificates. Some may think of the method as an extra protective layer to current Internet encryption protocols and it may
be used as such. In fact, the system operates on top of all IP layers and does not disturb current protocols in use by
Internet providers, including security ones. Anyway, it should be emphasized that the proposed method relies on security
created by physical noise. This way, its security level does not depend on advances in algorithms or computation. Also, the
proposed protocol does not need to be modified to follow improvements in communication's technology. Moreover, it works as a
stand-alone system.

 Random events of physical origin cannot be
deterministically predicted and sometimes are classified in classical or quantum
 events (See examples of differences between quantum and classical random walks in \cite{ChildsFarhiGutmann}).
 The point of view adopted here is that a recorded classical random event is
just the record of a single realization among all the possible quantum trajectories possible \cite{belavkin}. All of these
 distinct classifications are not relevant to the practical aspects to be discussed here. However, what should be emphasized
is that physical noise is completely different from pseudo noise generated in a deterministic process (e.g. hardware stream
ciphers) where despite any complexity introduced, the deterministic generation mechanism can be searched, eventually
discovered and used by the attacker.

\section{Signal encoding}

%\subsection{Mixing deterministic bits and noisy signals}

Before introducing the communication protocol to be used, one should
discuss the superposition of physical signals to deterministic
binary signals. Any signal transmitted over Internet is physically
prepared to be compatible with the channel being used. This way,
e.g., voltage levels $V_0$ and $V_1$ in a computer may represent
bits. These values may be understood as the simple encoding
\begin{eqnarray}
V^{(0)}\Rightarrow\left\{
\begin{array}{c}
V_0 \rightarrow \mbox{bit} \:0\\V_1 \rightarrow \mbox{bit}\:1 \end{array} \right.
\end{eqnarray}
Technical noise, e.g. electrical noise, in bit levels $V_0$ and $V_1$ are assumed low. Also, channel noise are assumed with
a modest level. Errors caused by these noises are assumed to be possibly corrected by classical error-correction codes.
Anyway, the end user is supposed to receive the bit sequence ${\bf X}$
 (prepared by a sequence of $V_0$ and $V_1$) as determined by the
sender.

 If one of these deterministic binary signals $x_j$ is repeated over the channel, e.g. $x_1=x$ and $x_2=x$, one has
the known property $x_1\oplus x_2=0$. This property has to be compared to cases where a non-negligible amount of physical
noise $n_j$ (in analog or a discrete  form) has been added to each emission. Writing $y_1=f_1(x_1,n_1)=f_1(x,n_1)$ and
$y_2=f_2(x_2,n_2)=f_2(x,n_2)$ one has $f(y_1,y_2)=$
%=f_?(n_1, n_2)=$
 neither $0$ or $1$ in general.
This {\em fundamental} difference from the former case where $x_1\oplus x_2=0$ emphasizes the uncontrollable effect of the
noise.

The $V^{(0)}$ encoding shown above allows binary values $V_0$ and $V_1$ to represent bits 0 and 1, respectively. These
values are assumed to be determined without ambiguity. Instead of this unique encoding  consider that {\em two} distinct
encodings can be used to represent bits 0 and 1: Either $V^{(0)}$ over which $x^{(0)}_0$ and $x^{(0)}_1$ represent the two
bits 0 and 1 respectively, or $V^{(1)}$, over which $x^{(1)}_1=x^{(0)}_0+\epsilon$ and $x^{(1)}_0=x^{(0)}_1+\epsilon$
($\epsilon \ll 1$) represent the two bits 1 or 0 (in a different order from the former assignment). These encodings
represent physical signals as, for example, phase signals.

Assume noiseless transmission signals but where noise $n_j$ has been introduced or added to each $j^{\mbox{\tiny th}}$ bit
sent (This is equivalent to noiseless signals in a noisy channel). Consider that the user does not know which encoding
$V^{(0)}$ or $V^{(1)}$ was used. With a noise level $n_{j}$ superposed to signals in $V^{(0)}$ or $V^{(1)}$ and if
$|x^{(0)}_0-x^{(1)}_0|\gg n_j \gg \epsilon$, one cannot distinguish between signals 0 and 1 in $V^{(0)}$ and
$V^{(1)}=V^{(0)}+\epsilon$ but one knows easily that a signal belongs either to the set ``$0 \:\:\mbox{in}\: V^{(0)}\:
\mbox{ or} \:1 \mbox{ in}\:  \:V^{(1)}$'' {\bf or} to the set ``$1 \:\mbox{ in}\: V^{(0)}\: \mbox{or}\: 0 \:\mbox{in}\:
V^{(1)}$''. Also note that once the encoding used is known, there is no difficulty to identify between $x_j$ and
$x_j+\epsilon$. In this case, it is straightforward to determine a bit 0 or 1 because values in a single encoding are widely
separated and, therefore, distinguishable. One may say that without information on the encoding used, the bit values cannot
be determined.

 Physical noise processes will be detailed ahead but this indistinguishability of the signals
 without basis information is the clue for A and B to share random bits over the Internet in a secure way.

Encryption methods with randomized ciphers have been proven to be secure, e.g.,  when the attacker's memory is limited
\cite{Maurer}. More recently, physical noise has been used both in free propagation and fiber-optics based systems using
$M$-ry levels \cite{Mry} for data encryption ($\alpha\eta$ systems) and have been analyzed ever since. See a recent
discussion in \cite{HirotaKurosawa}. The system proposed here is distinct from those $\alpha\eta$ systems but the idea of
covering information with physical noise underlies both logical structures. The proposed system is closer related to
 to the key distribution system presented in \cite{barbosaKey} but differs from it by the use of a deterministic
channel to carry noisy-recorded information (Some other aspects have been shown in \cite{otherquantph1} and
\cite{otherquantph2}) and the use of just two encryption bases \cite{BennettWiesner}. Use of the (ideally) deterministic
channel --or high intensity channels-- makes this system slower than the $\alpha\eta$ systems \cite{Mry} or
\cite{barbosaKey} due to the recording stage but avoids amplification problems related to those systems.

\section{Distribution protocol}

A brief description of protocol steps will be made, before a theoretic-security analysis that includes the system's
limitations is presented.
%It was said that if A and B start sharing a secret key sequence ${\bf K_0}$ beforehand they may
%end up with a secure fresh key sequence ${\bf K}$ much longer than ${\bf K_0}$ (${\bf K} \gg {\bf K_0}$).
Assume that shared sequence ${\bf K_0}$ gives encoding information, that is to say, which encoding ($V^{(0)}$ or $V^{(1)}$)
is being used at the $j^{\tiny \mbox{th}}$ emission. Assume that ${\bf K}_0=k_1^{(0)},k_2^{(0)},...$ has length $K_0$ and
that the user A has a physical random generator PhRG able to generate random bits {\em and} noise in continuous levels. A
generates a binary random sequence ${\bf K}_1=k_1^{(1)},k_2^{(1)},...k_{K_0}^{(1)}$ (say, binary voltage levels)  and a
sequence of $K_0$ noisy-signals $n$ (e.g., voltage levels in a continuum). The deterministic sequence (carrying recorded
noise) ${\bf Y}_1=k_1^{(0)}\oplus f_1( k_1^{(1)},n_1^{(1)}),k_2^{(0)}\oplus f_2(k_2^{(1)},n_2^{(1)}),...$ is then sent to B.
First, one has to see if B is able to extract the fresh sequence ${\bf K}_1$ from ${\bf Y}_1$:  B applies $f\left( {\bf Y}_1
,{\bf K_0}\right) \equiv f_1(k_1^{(1)},n_1^{(1)}),f_2(k_2^{(1)},n_2^{(1)}),...f_N(k_N^{(1)},n_N^{(1)})$. As B knows the
encoding used and the signals representing bits 0 or 1 in a given encoding are easily identifiable. B obtains $\:\:\:$
$f_1(k_1^{(1)},n_1^{(1)})\rightarrow k_1^{(1)},f_2(k_2^{(1)},n_2^{(1)})\rightarrow
k_2^{(1)},...f_N(k_N^{(1)},n_N^{(1)})\rightarrow k_N^{(1)}$. B then obtains the new random sequence ${\bf K}_1$ generated by
A.

Is the attacker also able to extract the same sequence ${\bf K}_1$?
Actually, this was a one-time pad with ${\bf K}_0$ with added noise
and, therefore, it is known that the attacker {\em cannot} obtain
${\bf K}_1$. The security problem arises for further exchanges of
random bits, e.g. if B wants to share further secret bits with A.

%\subsection{B sends a fresh random sequence to A}

Assume that B also has a physical random generator PhRG able to generate random bits and noise in continuous levels (One
could proceed with A being the sole one generating random signals but the problem is identical). B wants to send in a secure
way a freshly generated key sequence ${\bf K}_2=k_1^{(2)},k_1^{(2)},...k_{K_0}^{(2)}$ from his PhRG to A. B record the
signals ${\bf Y}_2=k_1^{(1)}\oplus f_1( k_1^{(2)},n_2^{(2)}),k_2^{(1)}\oplus f_2(k_2^{(2)},n_2^{(2)}),...$ and sends it to
A. As A knows ${\bf K}_1$  he(or she) applies ${\bf Y}_2 \oplus {\bf K_1}$ and extracts ${\bf K}_2$. A and B now share the
two new sequences ${\bf K}_1$ and ${\bf K}_2$. For speeding communication, even a simple rounding process to the nearest
encoding position would produce a simple binary output for the operation $f_j(k_j,n_j)$. The security of this process will
be shown after a presentation of the complete distribution protocol.

The simple description presented show a key distribution from A to B and from B to A, with the net result that A and B share
the fresh sequences ${\bf K}_1$ and ${\bf K}_2$. These steps can be seen as a first  distribution cycle.  A could again send
another fresh sequence ${\bf K}_3$ to B and so on. This repeated procedure provides A and B with sequences ${\bf K}_1,{\bf
K}_2,{\bf K}_3,{\bf K}_4,...$. This is the basic and simple key distribution protocol for the system.

A caveat should be made. Although the key sharing seems adequate to go without bounds, physical properties impose some
constraints and length limitations as discussed ahead.

\section{Physical encoding}

A and B use PhRGs to generate physical signals creating the random bits that define the key sequences ${\bf K}$ and the
continuous noise ${\bf n}$  necessary for the protocol. Being physical signals, precise variables have to discussed and the
noise source well characterized. Analog-to-digital interfaces will transform the physical signals onto binary sequences
adequate for Internet transmission protocols.
 Optical sources for the noise signals can be chosen for fast speeds. PhRGs have been
 discussed in the literature and even commercial ones are now
 starting to be available. Increasing operational speeds are expected. Without going into details one could divide the PhRG in
 two parts, one generating random binary signals and another
 providing noise in a continuous physical variable (e.g., phase of a light
 field). These two signals are detected, adequately formatted and can be
 added.

 Taking the phase of a light field as the physical variable of interest,
 one could assume laser light in a coherent state with average number of photons $\langle  n \rangle$
 within one coherence time ($\langle  n \rangle=|\alpha|^2\gg 1$)
 and phase $\phi$.
Phases $\phi=0$ could define the bit 0 while $\phi=\pi$ could define
the bit 1.

\begin{figure}
\centerline{\scalebox{0.3}{\includegraphics{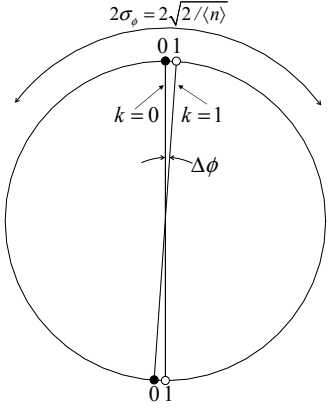}}} \caption{Bit positions in a phase sector with two possible
encodings (physical bases) defined by $k=0$ and $k=1$. Dark circles indicate positions for a bit 0 and open circles give
positions for a bit 1 on each encoding $k$. $\Delta \phi$ is the spacing between the two bases. $\sigma_{\phi}$ is the
standard phase deviation caused by phase fluctuations in a coherent light field.  $\langle n \rangle$ is adjusted so that
$\sigma_{\phi}$ covers close phase values.  $\Delta \phi$ should be kept $\ll \pi/2$. } \label{M2sector}
\end{figure}

A concrete image of a possible phase encoding with non-orthogonal states is seen in Fig. \ref{M2sector}. $k=0$ defines {\em
encoding} of 0 and 1 as phase values 0 and $\pi$, respectively. These values are widely separated and easily
distinguishable. This ease distinguishability will be  quantified ahead. Distinctly, $k=1$ defines encoding of 0 and 1 with
phase values $\pi+\Delta \phi$ and $0+\Delta \phi$, respectively. These bits are also easily distinguishable in $k=1$.
However, the poor distinguishability between 0s and 1s in distinct bases $k=0$ and $k=1$ is crucial for the proposed scheme;
this will be quantitatively explained.
 A bit 0 on this sector is encoded by one of the phases
\begin{eqnarray}
 \phi_{0k}=\left[
k \Delta \phi+\pi \frac{1-(-1)^{k}}{2} \right]\:\:, \:\:(k=0,1)\:\:,
\end{eqnarray}
while a bit 1 is encoded by
\begin{eqnarray}
 \phi_{1k}= \pi+\left[
k \Delta \phi+\pi \frac{1-(-1)^{k}}{2} \right]\:\:, \:\:(k=0,1)\:\:,
\end{eqnarray}

 It can be shown \cite{barbosaKey} (also shown ahead) that two non-orthogonal states with phases  $\phi_1$ and
 $\phi_2$ ($\Delta \phi_{12}=|\phi_1-\phi_2|\rightarrow 0$ and $\langle n \rangle \gg 1 $) overlap with (un-normalized) probability
\begin{eqnarray}
\label{pu} p_u \simeq e^{- (\Delta \phi_{12})^2 /2 \sigma_{
\phi}^2}\:\:,
\end{eqnarray}
where $\sigma_{\phi}=\sqrt{2/\langle n \rangle}$ is the standard deviation measure for the phase fluctuations $\Delta \phi$.
For distinguishable states, $p_u \rightarrow 0$ (no overlap) and for maximum indistinguishability $p_u=1$ (complete
overlap). With adequate formatting $\phi_1 -\phi_2$ gives the spacing $\epsilon$ ($\Delta \phi_{12} =\epsilon$) already
introduced. Eq. (\ref{pu}) with $\Delta \phi_{12}$ replaced by $\Delta \phi$   describes the probability for generic phase
fluctuations $\Delta \phi$ in a coherent state of constant amplitude $|\alpha|=\sqrt{\langle n \rangle}$.

The laser light intensity is adjusted by A (or B) such that $\sigma_{\phi} \gg \Delta \phi$. This guarantees that the
recorded information in the files to be sent over the open channel is in a condition such that the recorded light noise
makes the two close levels $\phi_1$ and $\phi_2$ highly indistinguishable to the attacker. In order to avoid the legitimate
user to confuse 0s and 1s in a {\em single} encoding, the light fluctuation should obey $\sigma_{\phi}\ll \pi/2$. These
conditions can be summarized as
\begin{eqnarray}
\label{condition}
 \frac{\pi}{2} \gg \sqrt{2/\langle n \rangle} \gg \Delta \phi
\:\:.
\end{eqnarray}
This shows that this key distribution system depends fundamentally
on physical aspects for security and not just on mathematical
complexity.

The separation between bits in the same encoding is easily carried under condition  $ \pi/2 \gg \sqrt{2/\langle n \rangle}$.
The condition $\sqrt{2/\langle n \rangle} \gg \Delta \phi$ implies that bits 0 (in encoding $k=0$) and 1 (in encoding $k=1$)
(upper position in Fig. \ref{M2sector}) cannot be easily identifiable and the same happens with sets of bit 1 (in encoding
$k=0$) and bit 0 (in encoding $k=1$) (lower position in Fig. \ref{M2sector}). However, for A, B and E, there are no
difficulty to identify that a sent signal is encoded by $k=0$ or $k=1$. One may therefore assume that physical signals
within the same encoding $k$ have negligible overlap. The signal distinguishability could also be studied assuming
non-negligible overlap between ``upper'' and ``lower'' states but results \cite{otherquantph2} are similar under the desired
conditions.

\subsection{Signal distinguishability}

The attacker does not know the encoding provided to A or B by their shared knowledge on the basis used. An answer to the
question ``What is the attacker's probability of error in bit identification without repeating a sent signal?'' depends on
the properties of the physical signals being used. Under the assumption that ``upper'' positions and ``lower'' positions in
Fig. \ref{M2sector} can be identified with high precision both by the legitimate users as well as by the attacker, this
question basically deals with distinguishability of the two close physical states in ``upper''or ``down'' positions.

 Binary
identification of two states has a general answer using information theory:
 The average probability of error in
identifying two states
 $| \psi_0\rangle$ and $| \psi_1\rangle$ is given by the Helstrom bound \cite{Helstrom}
\begin{eqnarray}
\label{Helstrom1}
 P_e=\frac{1}{2}\left[1-\sqrt{1-|\langle \psi_0|
\psi_1\rangle|^2}\: \right]\:\:.
\end{eqnarray}
Here $| \psi_0\rangle$ and $| \psi_1\rangle$ are coherent states of light $|\psi \rangle$ \cite{glauber} with same amplitude
but distinct phases
\begin{eqnarray}
| \psi \rangle =|\alpha \rangle=| |\alpha| e^{-i\phi}\rangle=e^{-\frac{1}{2}|\alpha|^2} \sum_n \frac{\alpha^n}{\sqrt{n!}}| n
\rangle \:, \end{eqnarray} defined at the PhRG. $| \psi_0\rangle$ define states in encoding $k=0$, where bits 0 and 1 are
given by
\begin{eqnarray}
| \psi_0\rangle=\left\{
\begin{array}{c}
\hspace{3mm}| \alpha  \rangle, \:\:\:\mbox{for bit}\:\:\: 0,\:\:\:\mbox{and} \\
| -\alpha  \rangle, \:\:\:\mbox{for bit}\:\:\: 1\:\:,\end{array}
\right.
\end{eqnarray}
$| \psi_1\rangle$ define states in encoding $k=1$, where bits 1 and 0 are given by
\begin{eqnarray}
| \psi_1\rangle=\left\{
\begin{array}{c}
| |\alpha |e^{-i \frac{\Delta \phi}{2}} \rangle, \:\:\:\mbox{for bit}\:\:\: 1,\:\:\:\mbox{and} \\
|  |\alpha| e^{-i \left( \frac{\Delta \phi}{2}+\pi \right)}
\rangle, \:\:\:\mbox{for bit}\:\:\: 0\:\:,\end{array} \right.
\end{eqnarray}
where $|\phi_{k=0} - \phi_{k=1}|=\Delta \phi$. $|\langle \psi_0| \psi_1\rangle|^2$ is calculated in a straightforward way
and gives
\begin{eqnarray}
|\langle \psi_0| \psi_1\rangle|^2=e^{-2 \langle n \rangle \left[ 1-\cos \frac{\Delta \phi}{2}\right]}\:\:.
\end{eqnarray}
 For $\langle n \rangle \gg 1$ and
$\Delta \phi\ll 1$,
\begin{eqnarray}
|\langle \psi_0| \psi_1\rangle|^2\simeq e^{- \frac{\langle n \rangle }{4}\Delta \phi^2} \equiv e^{- \Delta \phi^2/\left( 2
\sigma_\phi^2\right)}\:\:.
\end{eqnarray}

One should remind that in the proposed system the measuring procedure is defined by the users A and B and {\em no} physical
attack can be launched by E can improve the deterministic signals that were already available to him(her). Thus, her
knowledge on the signals cannot be increased by measurement techniques.

\section{Information leak and length limitation}

One should observe that each random bit defining the key sequence is once sent as a message by A (or B) and then resent as a
key (encoding information) from B (or A) to A (or B). In a deterministic encryption this will lead straightforwardly to a
breaking of the security. The noisy signals modify this situation dramatically: In both emissions, noise is superposed to
the signals. In general, repetitions of coherent signal imply that a better resolution may be achieved that is proportional
to the number of repetitions $r$. This improvement in resolution is equivalent to a single measurement with a signal
$r\:\:\times$ more intense. To take into account the single repetition demanded by the protocol $\langle n \rangle$ is
replaced
 by $2\langle n \rangle$ in $|\langle \psi_0| \psi_1\rangle|^2$. In other words, the protection level will then be
 considered for signal levels twice stronger than the one currently used. The final probability of error  results
\begin{eqnarray}
P_e=\frac{1}{2}\left[1-\sqrt{1-e^{-\frac{\langle n \rangle}{2} \Delta \phi^2}} \:\right]\:\:.
\end{eqnarray}

This error probability  can be used to derive some of the proposed system's limitations. The attacker's probability of
success $P_s\:( =1-P_e)$ to obtain the basis used in a single emission may be used to compare with the a-priori starting
entropy $H_{k}$ of the encoding that carries one bit of the message to be sent (a random bit). If the attacker knows the
encoding, the bit will also be known, with the same probability $\rightarrow 1$ as the legitimate user.
\begin{eqnarray}
H_{k,\small \mbox{bit}}=-p_0 \log p_0- p_1 \log p_1=1\:\:,
\end{eqnarray}
where $p_0$ and $p_1$ are the a-priori probabilities for each encoding $k$ ($p_0=p_1=1/2$) given by the PhRG. The entropy
defined by success events is $H_s=-P_s \log P_s$.  The entropy variation $\Delta H=H_{k,\small \mbox{bit}}-H_s$
statistically obtained --or leaked from bit measurements-- show the statistical information acquired by the attacker with
respect to the a-priori starting entropy:
\begin{eqnarray}
\Delta H_{k}=\left( H_{k,\small \mbox{bit}}-H_s \right)\:\:.
\end{eqnarray}
\begin{figure}
\centerline{\scalebox{0.6}{\includegraphics{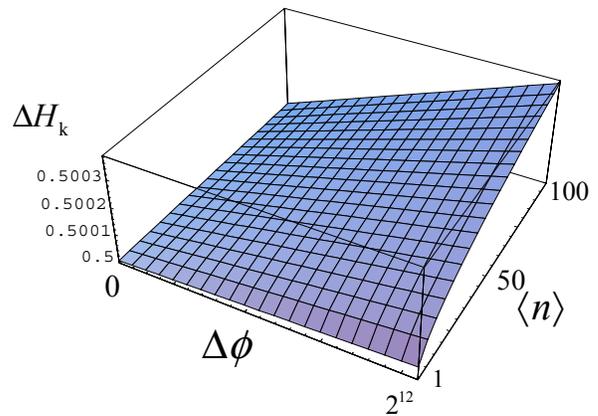}}} \caption{ $\Delta H_{k}$ as a function of $\langle n \rangle$ and
$\Delta \phi$.} \label{delH}
\end{figure}

Fig. \ref{delH} shows $\Delta H_{k}$ for some values of $\langle n \rangle$ and $\Delta \phi$. Value $\Delta H_{k}=1/2$ is
the limiting case where the two bases cannot be distinguished. $\Delta H_{k}$ deviations from this limiting value of $1/2$
indicates that some amount of information on the basis used may potentially be leaking to the attacker. However, it is clear
that the attacker cannot obtain bit-by-bit the encoding used.

\subsection{Length limitation}
In order to be possible to obtain {\em statistically} a good amount of information on a {\em single} encoding used, $L$ bits
have to be transmitted. $L$ thus establishes the length limitation for bits exchanged starting from ${\bf K}_0$. It will be
defined by
\begin{eqnarray}
\label{length}
 L \times \left( \Delta H_{k}-\frac{1}{2}\right) = 1\:\:.
\end{eqnarray}
\begin{figure}
\centerline{\scalebox{0.45}{\includegraphics{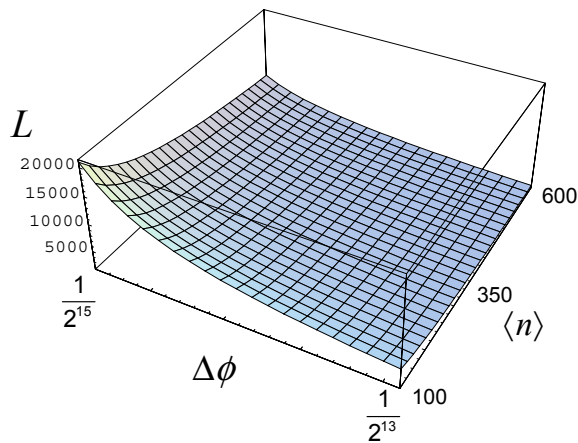}}}
\caption{Estimates for the minimum length of bits $L$ exchanged
between A and B that could give {\em one} bit of information about
the bases used to the attacker.} \label{leak}
\end{figure}
Fig. \ref{leak} shows estimates for $L$ for a range of values $\langle n \rangle$ and $\Delta \phi$ satisfying $\label{L} L
\times \left( \Delta H_{k}-\frac{1}{2}\right) = 1\:\:$($\Delta \phi$ is given in powers of 2, indicating bit resolution for
analog-to-digital converters). It should be emphasized that $\langle n \rangle$ is the {\em mesoscopic} average photon
number in the PhRG while an optical signal in the transmission channel can be carried by very intense light--the
deterministic signal.

It is assumed that error correction codes can correct for technical
errors in the transmission/reception steps for the legitimate users.
The leak estimate given by Eq. (\ref{length}) do not imply that the
information actually has leaked to the attacker.
 However, for security reasons, one takes for granted that this
deviation indicate a statistical fraction of bits acquired by the attacker. A probability measure corresponds to condition
(\ref{length}): $p(1)=1/L=\left( \Delta H_{k}-\frac{1}{2}\right)$, expressing that  one bit among L bits may have been
statistically compromised.

Privacy amplification procedures can be applied to the shared bits in order to reduce this hypothetical information gained
by the attacker to negligible levels \cite{Wolf}. These procedures are beyond the purposes of the present discussion but one
can easily accept that A and B may discard a similar fraction of bits to statistically reduce the amount of information
potentially leaked. Reducing this fraction of bits after a succession of bits are exchanged between A and B implies, e.g.,
that the number of bits to be exchanged will decrease at every emission. Eventually, a new shared key ${\bf K}_0$ has to
start the process again to make the system secure. Nevertheless, the starting key length $K_0$ was boosted in a secure way.
Without further procedures, the physical noise allowed  ${\bf K} \gg 10^3 {\bf K}_0$, a substantial improvement over the
classical one-time pad factor of 1. One may still argue that the ultimate security relies on ${\bf K}_0$'s because if ${\bf
K}_0$ is known no secret will exist for the attacker. This is also true but does not invalidate the practical aspect of the
system.  ${\bf K}_0$ length can even be made sufficiently long to frustrate any brute-force attack at any stage of
technology. Therefore, the combination of physical noise and complexity makes this noisy-one-time pad practical for Internet
uses.

\section{Frustrating a-posteriori known-plaintext attacks}

\subsection{Known-plaintext attack}

Although the security of the process has been demonstrated, one should also point to a  fragility of the system that has to
be avoided when A and B are encrypting messages ${\bf X}$ between them. As it was shown, knowledge of one sequence of random
bits lead to the knowledge of the following sequence for A and B. This makes the system vulnerable to know-plaintext attacks
in the following way:
 E has a perfect record of sequences
${\bf Y}_1$ and ${\bf Y}_2$ and tries to recover any key sequence from them, ${\bf K}_2$, ${\bf K}_1$ or ${\bf K}_0$. E will
wait until A and B uses these sequences for encryption before trying to brake the system. A and B will encrypt a message
using a new shared sequence, ${\bf K}_1$ or ${\bf K}_2$. This message could be a plaintext, say ${\bf X}=x_1,x_2,...x_{K_0}$
{\em known} to the attacker. Encrypting this message with say ${\bf K}_1$ in a noiseless way, gives ${\bf Y}=x_1 \oplus
k_1^{(1)},x_2\oplus k_2^{(1)},...x_{K_0}\oplus k_{K_0}^{(1)}$. Performing the operation ${\bf Y}\oplus {\bf X}$, E obtains
${\bf K}_1$. The chain dependence of ${\bf K}_j$ on ${\bf K}_{j-1}$ allows E to find successive keys. Even addition of noise
to the encrypted file does not eliminate this fragility, because the attacker can use his/her knowledge of ${\bf X}$ --as
the key-- to obtain ${\bf K}$--as a message. The situation is {\em symmetric} between B or the attacker: one that knows the
key (${\bf X}$ for E, and ${\bf K}$ for B) obtains the desired message (${\bf K}$ for E, and ${\bf X}$ for B).

This kind of attack can be frustrated to the attacker with a simple strategy as explained ahead.  In general, random
generation processes are attractive to attackers. Even  physical components (e.g. PhRG) are targets for attackers that may
try to substitute a true random sequence by pseudo-random bits generated by a seed key under his/her control. Electronic
components can also be inserted to perform this task replacing the original generator; electric or electromagnetic signal
may induce sequences for the attacker and so on. While these can be controlled by simple equipment surveillance, the
known-plaintext attack is more subtle. A may not know, e.g., that some information to be transmitted to B is known to the
attacker.

\subsection{Frustrating the known-plaintext attack}

This attack can be avoided by shuffling (permutation operations) the random bit sequence being transmitted from A to B (or B
to A) followed by a re-shuffling by B (or A). A particular shuffling function could be chosen among members of a family of
one-way functions by use of a short sequence of shared secret random bits. An even simpler way (or less costly) is to use
the short sequence of bits to choose one among a list of pre-recorded permutations, what speeds the processing time. The
shuffling function used changes from block to block of sequences exchanged. This creates a non-invertible structure for the
attacker.

Although the number of bits $n_s$ necessary to select one among $K_0!$ permutations in $K_0$ bits has a too high cost,
 a reduced number of permutations $K_0!/d$ within $K_0!$ can be chosen to provide the pre-recorded list and still provide
 a negligible chance for the attacker to obtain a particular choice among $K_0!/d$. With this reduced list, the number of
 bits $n_b$ necessary to assign a particular permutation is given by $K_0!/d=2^{n_b}$. At the same time the probability $p_d$
for the attacker to find this particular permutation choice is $p_d=1/(K_0!/d)$.

The a-posteriori known-plaintext attack can then be frustrated by the shuffled sequence. The random bit sequence used in
this process is also protected by the encryption method with the added noise.

\section{Message authentication} Encryption performed by one user with the shared random bits can only be decrypted by the
other legitimate user and the message obtained can therefore be understood as ``authentic''. However, it may happen that
particular messages need explicit message authentication without the need to decrypt. Fortunately, this can be done with a
modest use of shared secret random bits. For example, Ref.~\cite{HMAC} describes a message authentication code (MAC) where
{\em one} key with $k$ shared secret bits encode message blocks and generate a tag $T$. No decryption is needed, the
receiver applies the shared key to the received data stream to generate the tag $T^{\prime}$. Authenticity is given by
$T^{\prime}=T$.

\section{Conclusions}

As a conclusion, it has been shown that Internet users will succeed in generating and sharing, in a fast way, a large number
of secret keys to be used in bit-by-bit encryption (one-time pad).  They have to start from a shared secret sequence of
random bits obtained from a {\em physical} random generator. The physical noise in the signals openly transmitted is set to
hide the random bits sent. No intrusion detection method is necessary. Privacy amplification protocols eliminate any
fraction of information that may have eventually obtained by the attacker. As the security is not only based on mathematical
complexities but depend on physical noise, scientific or technological advances will not harm this system. This is then very
different from systems that would rely entirely,  say, on the current lack of efficient algorithms to factor large numbers
into their primes. The system is also  secure against a posteriori known-plaintext attacks on the key. It was then shown
that by sharing secure secret key sequences, a practical bit-by-bit encryption over the Internet can be implemented.  \\\\
$^*$E-mail:$\:\:$GeraldoABarbosa@hotmail.com

\end{document}